\documentclass[12pt,nofootinbib, preprintnumbers, superscriptaddress]{revtex4-1}

\usepackage{amsmath,amssymb,amscd,simplewick}
\usepackage{listings}
\usepackage{dsfont}
\usepackage{slashed}
\usepackage{color}
\usepackage{ulem}

\usepackage{graphicx}
\usepackage{epstopdf}
\usepackage{subfigure}
\usepackage{epsfig}

\usepackage{xcolor}
\usepackage[colorlinks=true,
            linkcolor=blue,
            urlcolor=blue,
            citecolor=green,
            bookmarks=true,
            bookmarksnumbered=true,
            breaklinks=true,
            pdfpagemode=Fullscreen,
            pdfstartview=FitBH]{hyperref}

\begin{document}

\title{A translational flavor symmetry in the mass terms of \\ Dirac and
Majorana fermions}
\author{Zhi-zhong Xing}
\email{xingzz@ihep.ac.cn}
\affiliation{Institute of High Energy Physics, Chinese Academy of Sciences, Beijing 100049, China}
\affiliation{School of Physical Sciences, University of Chinese Academy of Sciences, Beijing 100049, China}
\affiliation{Center of High Energy Physics, Peking University, Beijing 100871, China}

\begin{abstract}
\vspace{2cm}
Requiring the effective mass term for a category of fundamental Dirac or Majorana
fermions of the same electric charge to be invariant under the translational
transformations $\psi^{}_{\alpha \rm L (R)} \to \psi^{}_{\alpha \rm L (R)} +
n^{}_{\alpha} z^{}_{\psi \rm L(R)}$ in the flavor space, where $n^{}_\alpha$ and
$z^{}_{\psi \rm L(R)}$ stand respectively for the flavor-dependent complex numbers
and a constant spinor field anticommuting with the fermion fields,
we show that $n^{}_\alpha$ can be identified as
the elements $U^{}_{\alpha i}$ in the $i$-th column of the unitary matrix $U$ used
to diagonalize the corresponding Hermitian or symmetric fermion mass matrix $M^{}_\psi$,
and $m^{}_i = 0$ holds accordingly. We find that the reverse is also true. Now
that the mass spectra of charged leptons, up- and down-type quarks are all strongly
hierarchical and current experimental data allow the lightest neutrino to be
massless, we argue that the zero mass limit for the first-family fermions and
the translational flavor symmetry behind it should be a natural starting
point for building viable fermion mass models.
\end{abstract}

\maketitle

\def\thefootnote{\arabic{footnote}}
\setcounter{footnote}{0}

\newpage

\section{Introduction}

The standard model (SM) of particle physics has proved to be a huge
success, but its flavor sector remains unsatisfactory in the sense that all the flavor
parameters are theoretically undetermined. Going beyond the SM, one has to introduce
some more free parameters in connection with the massive neutrinos. One way out of
this situation is to determine or constrain the flavor structures of leptons
and quarks with the help of certain proper family symmetries,
such that some testable predictions for fermion masses and flavor mixing
parameters (or their correlations) can be achieved. So far a lot of efforts
have been made along this line of thought toward building phenomenologically
viable models \cite{Xing:2019vks,Feruglio:2019ybq},
but the true flavor dynamics is still unclear. At the present stage
a logical candidate for the underlying flavor symmetry should at least help
interpret some salient features of the observed fermion mass spectra and flavor
mixing patterns.

In 2006 Friedberg and Lee put forward a novel idea to constrain the
flavor texture of three Dirac neutrinos \cite{Friedberg:2006it}.
In the basis of a diagonal charged-lepton mass matrix, they required
the Dirac neutrino mass term ${\cal L}^{}_\nu$ to keep unchanged
under the transformation $\nu^{}_\alpha \to \nu^{}_\alpha + z^{}_\nu$
(for $\alpha = e, \mu, \tau$), where $z^{}_\nu$ is a constant spinor field
anticommutating with the neutrino fields $\nu^{}_\alpha$ \cite{Volkov:1973ix}.
The constraint of this {\it translational} symmetry of ${\cal L}^{}_\nu$ on the neutrino
mass matrix $M^{}_\nu$ is so strong that $\det\left(M^{}_\nu\right) = 0$ holds, implying
that one of the neutrino masses $m^{}_i$ (for $i=1,2,3$) must vanish
(i.e., one of the neutrinos is a Goldstone-like fermion
\cite{Volkov:1973ix,deWit:1975xci}). A very instructive neutrino mixing pattern
$U = U^{}_{\rm TBM} O^{}_{13}$ in correspondence to $m^{}_2 =0$ can then be
obtained \cite{Friedberg:2006it,Xing:2006xa,Friedberg:2007ba,Friedberg:2007uk,Araki:2010zb},
where $U^{}_{\rm TBM}$ stands for the well-known ``tribimaximal" flavor mixing pattern
\cite{Harrison:2002er,Xing:2002sw,He:2003rm} and $O^{}_{13}$ is a unitary
rotation matrix in the complex $(1,3)$ plane. In this scenario, however, a proper
symmetry breaking term has to be introduced to assure $m^{}_2 > m^{}_1$
as indicated by current neutrino oscillation data \cite{Zyla:2020zbs}.

In fact, as early as in 1979, 't Hooft had imposed a quite similar translational
displacement $\phi(x) \to \phi(x) + \Lambda$ on the Lagrangian ${\cal L}^{}_\phi$
for a renormalizable scalar field theory, where $\Lambda$ is a constant scalar
field commutating with $\phi$ \cite{tHooft:1979rat}.
It turns out that both the mass and the self-coupling parameter of $\phi$
have to vanish in order to guarantee the invariance of ${\cal L}^{}_\phi$, which
can be referred to as a Goldstone-type symmetry
\cite{Goldstone:1961eq,Goldstone:1962es}, under the above transformation.
So one generally expects that the translational symmetry
of an effective Lagrangian may provide a simple and natural way to understand why
the mass of a fermion or boson in this physical system is vanishing or vanishingly small
\footnote{If there were no flavor mixing in the lepton or quark sector, the
flavor and mass eigenstates of a fermion would be identical with each other.
In this case the kinetic energy and mass terms of a fermion field, which does
not involve any self-interactions, would be closely analogous to
those of a scalar field.},
although its deep meaning remains a puzzle at present.

In this paper we point out that the translational transformation
$\nu^{}_i \to \nu^{}_i + z^{}_\nu$ for a massless neutrino field $\nu^{}_i$
(i.e., $m^{}_i = 0$ for either $i=1$ or $i=3$) is equivalent to the translational
transformation $\nu^{}_\alpha \to \nu^{}_\alpha + U^{}_{\alpha i} z^{}_\nu$
in the flavor space, where $U^{}_{\alpha i}$ (for $\alpha = e, \mu, \tau$) denote
the corresponding neutrino flavor mixing matrix elements and $z^{}_\nu$
represents a constant spinor field anticommuting with the neutrino fields.
Then we are going to show that the effective mass terms of Dirac or Majorana
fermions may all have a kind of translational symmetry under the discrete shifts
$\psi^{}_{\alpha \rm L (R)} \to \psi^{}_{\alpha \rm L (R)} + n^{}_{\alpha}
z^{}_{\psi \rm L(R)}$ for a constant spinor field $z^{}_{\psi \rm L (R)}$
in the flavor space, if and only if
$m^{}_i = 0$ holds and $n^{}_\alpha = U^{}_{\alpha i}$
are the elements in the $i$-column of the unitary matrix $U$ used to diagonalize
the corresponding Hermitian or symmetric fermion mass matrix $M^{}_\psi$.
We find that the reverse is also true. Given the very facts that all the fundamental
fermions of the same nonzero electric charge have a rather strong mass hierarchy and
current neutrino oscillation data allow the lightest neutrino to be (almost)
massless, the zero mass limit for the first-family fermions and the translational
symmetry behind it can serve as a very natural starting point for building viable
fermion mass models toward understanding the observed patterns of both the
Cabibbo-Kobayashi-Maskawa (CKM) quark flavor mixing matrix
\cite{Cabibbo:1963yz,Kobayashi:1973fv} and the Pontecorvo-Maki-Nakagawa-Sakata
(PMNS) lepton flavor mixing matrix \cite{Pontecorvo:1957cp,Maki:1962mu,Pontecorvo:1967fh}.

\section{Translation of a massless neutrino field}

It is well known that the massless photon travels at the speed of light in free
space and its electromagnetic field obeys the equation of motion
$\square {\bf A} = 0$. This equation is invariant under the translational
transformation ${\bf A} \to {\bf A} + {\bf A}^{}_0$, where ${\bf A}^{}_0$ denotes
a constant vector field commuting with ${\bf A}$. Similarly, a massless neutrino
also travels at the speed of light in free space and its
field $\nu^{}_i$ (for $i = 1$ or $3$) satisfies the Dirac
equation ${\rm i}\gamma^\mu \partial^{}_\mu \nu^{}_i = 0$, which is invariant
under the translational transformation
\begin{eqnarray}
\nu^{}_i \to \nu^{}_i + z^{}_\nu \; ,
\end{eqnarray}
where $z^{}_\nu$ is a constant spinor field anticommuting with the neutrino
fields. This observation means that such a translation of the massless neutrino
field is consistent with Einstein's principle of constancy of light velocity
for all the inertial reference systems in vacuum.

Although current neutrino oscillation data {\it do} allow the existence of
a massless neutrino species, the other two neutrino species must be massive.
In this case the massless neutrino field and its two massive counterparts
can form three quantum superposition states --- the neutrino flavor
eigenstates, which directly participate in the standard weak interactions. This kind of
mismatch between the flavor and mass eigenstates of three neutrinos is described
by a $3\times 3$ unitary matrix $U$ as follows:
\begin{eqnarray}
\left(\begin{matrix} \nu^{}_e \cr \nu^{}_\mu \cr \nu^{}_\tau \cr \end{matrix}
\right)_{\hspace{-0.07cm} \rm L}
= \left(\begin{matrix} U^{}_{e 1} & U^{}_{e 2} & U^{}_{e 3} \cr
U^{}_{\mu 1} & U^{}_{\mu 2} & U^{}_{\mu 3} \cr
U^{}_{\tau 1} & U^{}_{\tau 2} & U^{}_{\tau 3} \cr \end{matrix} \right)
\left(\begin{matrix} \nu^{}_1 \cr \nu^{}_2 \cr \nu^{}_3 \cr \end{matrix}
\right)_{\hspace{-0.07cm} \rm L} \; .
\end{eqnarray}
As a consequence, the translational transformation made for the massless neutrino
field $\nu^{}_i$ with $m^{}_i =0$ in Eq. (1) requires that the three neutrino flavor
states $\nu^{}_\alpha$ (for $\alpha = e, \mu, \tau$) transform in the following way:
\begin{eqnarray}
\nu^{}_{\alpha \rm L} \to \nu^{}_{\alpha \rm L} + U^{}_{\alpha i} z^{}_\nu \; .
\end{eqnarray}
So the aforementioned Friedberg-Lee transformation \cite{Friedberg:2006it} is
equivalent to taking $U^{}_{e 2} = U^{}_{\mu 2} = U^{}_{\tau 2} = 1/\sqrt{3}$
in correspondence to $m^{}_2 =0$.

Now that the translational transformation described by Eq. (3) in the flavor space
is equivalent to the translation of a massless neutrino field described by Eq. (1),
one should be able to show that $m^{}_i =0$ must hold if the effective Majorana
(or Dirac) neutrino mass term keeps invariant under the transformation in Eq. (3).
In the subsequent sections we are going to demonstrate that this is really the case.

\section{Majorana neutrinos}

First of all, let us consider the effective mass term of the three known neutrinos
(i.e., $\nu^{}_e$, $\nu^{}_\mu$ and $\nu^{}_\tau$) by
assuming that they have the Majorana nature
\footnote{Of course, an effective Majorana mass term for the right-handed neutrino fields
in the canonical seesaw mechanism \cite{Minkowski:1977sc,Yanagida:1979as,GellMann:1980vs,
Glashow:1979nm,Mohapatra:1979ia} can be discussed in an analogous way.}
\begin{eqnarray}
-{\cal L}^{}_{\rm M} = \frac{1}{2} \sum_\alpha\sum_\beta \left[\overline{\nu^{}_{\alpha \rm L}}
\hspace{0.1cm} \langle m\rangle^{}_{\alpha\beta} \left(\nu^{}_{\beta \rm L}\right)^c \right]
+ {\rm h.c.} \; ,
\end{eqnarray}
where $\alpha$ and $\beta$ run over the flavor indices $e$, $\mu$ and $\tau$,
``$c$" denotes the charge conjugation, and
$\langle m\rangle^{}_{\alpha\beta} = \langle m\rangle^{}_{\beta \alpha}$
are the elements of the $3\times 3$ Majorana neutrino mass matrix
$M^{}_\nu$. Since $M^{}_\nu$ can be diagonalized via the unitary transformation
$U^\dagger M^{}_\nu U^* = {\rm diag}\{m^{}_1, m^{}_2, m^{}_3\}$, we have
the expressions
\begin{eqnarray}
\langle m\rangle^{}_{\alpha\beta} = \sum_i \left(m^{}_i U^{}_{\alpha i} U^{}_{\beta i}
\right) \; ,
\end{eqnarray}
for $i=1,2$ and $3$. The unitarity of $U$ allows us to prove
\begin{eqnarray}
&& \sum_\alpha \left[U^{*}_{\alpha j} \langle m\rangle^{}_{\alpha\beta} \right]
= m^{}_j U^{}_{\beta j} \; ,
\nonumber \\
&& \sum_\beta \left[\langle m\rangle^{}_{\alpha\beta} U^{*}_{\beta j} \right]
= m^{}_j U^{}_{\alpha j} \; ,
\nonumber \\
&& \sum_\alpha \sum_\beta \left[U^{*}_{\alpha j} \langle m\rangle^{}_{\alpha\beta}
U^{*}_{\beta j} \right] = m^{}_j \; ,
\end{eqnarray}
which are essentially equivalent to one another and all proportional to the
neutrino mass $m^{}_j$ (for $j=1,2$ or $3$). Following Eq. (3),
now we make the same translational transformation for the
left-handed neutrino fields in the flavor space
\footnote{Different from Ref. \cite{Friedberg:2006it} and some other references,
where only the Dirac neutrinos
and the simplest flavor-independent transformation $\nu^{}_\alpha \to \nu^{}_\alpha + z$
are taken into account, here we have considered a more generic and flavor-dependent
transformation for the Majorana neutrino fields. This nontrivial treatment will allow
us to establish a direct and thus more transparent connection between the vanishing
neutrino mass and the corresponding neutrino mixing matrix elements.},
\begin{eqnarray}
\nu^{}_{\alpha \rm L} \to \nu^{}_{\alpha \rm L} + U^{}_{\alpha j} z^{}_\nu \; ,
\end{eqnarray}
Then ${\cal L}^{}_{\rm M}$ becomes
\begin{eqnarray}
-{\cal L}^\prime_{\rm M} = -{\cal L}^{}_{\rm M} + \frac{1}{2} m^{}_j \left[
\overline{z^{}_\nu} \hspace{0.05cm} z^c_\nu +
\sum_\alpha \left[U^{}_{\alpha j} \hspace{0.05cm}\overline{\nu^{}_{\alpha \rm L}}
\right] z^c_\nu + \overline{z^{}_\nu} \sum_\beta \left[U^{}_{\beta j}
\left(\nu^{}_{\beta \rm L}\right)^c\right]\right] \; .
\end{eqnarray}
It becomes clear that ${\cal L}^\prime_{\rm M} = {\cal L}^{}_{\rm M}$ will hold under
the above transformation if and only if $m^{}_j = 0$ holds. Namely, one of the
three neutrinos must be massless if the effective Majorana neutrino mass term
${\cal L}^{}_{\rm M}$ keeps invariant under the discrete shifts of $\nu^{}_{\alpha \rm L}$
made in Eq.~(7), which helps provide a novel link between the two sides of one coin (i.e.,
the mass and flavor mixing issues of the Majorana neutrinos). {\it The point is that
the three flavor-dependent coefficients $U^{}_{\alpha j}$ of $z^{}_\nu$
constitute the $j$-th column of the unitary matrix $U$ used to diagonalize $M^{}_\nu$,
which corresponds to $m^{}_j =0$}. In comparison, most of the popular global discrete
flavor symmetries can help predict very specific neutrino mixing patterns but leave the
neutrino mass spectrum unconstrained \cite{Altarelli:2010gt,Ishimori:2010au,King:2013eh}.

Given the fact of $m^{}_2 > m^{}_1$, one is left with either
$m^{}_1 =0$ (normal ordering) or $m^{}_3 =0$ (inverted ordering) for the neutrino
mass spectrum. In either case the other two neutrino masses can be determined
by inputting the experimental values of two independent neutrino mass-squared
differences, and only a single nontrivial Majorana CP phase of $U$ survives
\cite{Xing:2020ald}.

In the basis of a diagonal charged-lepton mass matrix, the unitary matrix $U$ appearing
in Eqs.~(5)---(8) is just the PMNS matrix $U^{}_{\rm PMNS}$ which link
the neutrino mass eigenstates $\nu^{}_i$ (for $i=1,2,3$) to the neutrino flavor
eigenstates $\nu^{}_\alpha$ (for $\alpha = e, \mu, \tau$). Focusing on the
possibility of $m^{}_1 = 0$, one may adopt the original Kobayashi-Maskawa (KM)
parametrization \cite{Kobayashi:1973fv} for $U^{}_{\rm PMNS}$ so as to make
the expressions of $U^{}_{\alpha 1}$ in the first column of $U^{}_{\rm PMNS}$
as simple as possible. Explicitly,
\begin{eqnarray}
U^{}_{\rm PMNS} = \left(\begin{matrix} c^{}_1 & s^{}_1 c^{}_3 & s^{}_1 \hat{s}^{*}_3 \cr
-s^{}_1 c^{}_2 & c^{}_1 c^{}_2 c^{}_3 + s^{}_2 \hat{s}^{}_3 &
c^{}_1 c^{}_2 \hat{s}^{*}_3 - s^{}_2 c^{}_3 \cr
-s^{}_1 s^{}_2 & c^{}_1 s^{}_2 c^{}_3 - c^{}_2 \hat{s}^{}_3 &
c^{}_1 s^{}_2 \hat{s}^{*}_3 + c^{}_2 c^{}_3 \cr \end{matrix} \right)
P^{}_\nu \;  \hspace{0.4cm}
\end{eqnarray}
with the definitions $c^{}_i \equiv \cos\theta^{}_i$, $s^{}_i \equiv \sin\theta^{}_i$,
$\hat{s}^{}_3 \equiv s^{}_3 e^{{\rm i}\phi}$ and
$P^{}_\nu \equiv {\rm diag}\{1 , e^{{\rm i}\sigma} , 1\}$, where
$\phi$ and $\sigma$ denote the Dirac and Majorana CP-violating phases, respectively.
Provided $U^{}_{e 1} = 2/\sqrt{6}$ and $U^{}_{\mu 1} = U^{}_{\tau 1}
=-1/\sqrt{6}$ are assumed (i.e., $\theta^{}_1 = \arcsin(1/\sqrt{3})
\simeq 35.3^\circ$ and $\theta^{}_2 = 45^\circ$ are taken \cite{Zhao:2015bza}),
for example, one will obtain
\begin{eqnarray}
U^{}_{\rm PMNS} = \frac{1}{\sqrt 6} \left(\begin{matrix} 2 & \sqrt{2} & 0 \cr
-1 & \sqrt{2} & -\sqrt{3} \cr
-1 & \sqrt{2} & \sqrt{3} \cr \end{matrix} \right)
\left(\begin{matrix} 1 & 0 & 0 \cr
0 & c^{}_3 & \hat{s}^{*}_3 \cr
0 & - \hat{s}^{}_3 & c^{}_3 \cr \end{matrix} \right)
P^{}_\nu \; ; \hspace{0.3cm}
\end{eqnarray}
namely, $U^{}_{\rm PMNS} = U^{}_{\rm TBM} O^{}_{23} P^{}_\nu$ with $O^{}_{23}$ being a
unitary rotation matrix in the complex $(2,3)$ plane. This simple flavor mixing pattern,
which was first proposed in 2006 \cite{Xing:2006ms,Lam:2006wm}, remains favored
in today's neutrino phenomenology. A combination of Eqs.~(5) and (10) allows us to
reconstruct the Majorana neutrino mass matrix in the chosen basis:
\begin{eqnarray}
M^{}_\nu = \left(\begin{matrix} a & a & a \cr
a & a & a \cr
a & a & a \cr \end{matrix} \right) +
\left(\begin{matrix} 0 & c & -c \cr
c & b + 2c & -b \cr
-c & -b & b - 2c \cr \end{matrix} \right) \; ,
\end{eqnarray}
in which $a \equiv \left( \overline{m}^{}_2 c^2_3 + m^{}_3 \hat{s}^{*2}_3 \right)/3$,
$b \equiv \left( \overline{m}^{}_2 \hat{s}^{2}_3 + m^{}_3 c^2_3 \right)/2$ and
$c \equiv \left( \overline{m}^{}_2 \hat{s}^{}_3 - m^{}_3 \hat{s}^{*}_3
\right) c^{}_3 /\sqrt{6}$ with $\overline{m}^{}_2 \equiv m^{}_2 e^{2{\rm i}\sigma}$
are defined. The simple structure of $M^{}_\nu$ depends on the simple choice of
$U^{}_{\alpha 1}$ and is suggestive of certain simple flavor symmetries which can be
used for the explicit model building \cite{Lam:2006wm}.
Note that in this example the PMNS matrix $U^{}_{\rm PMNS}$
only possesses a partial $\mu$-$\tau$ permutation symmetry characterized by
$U^{}_{\mu 1} = U^{}_{\tau 1}$ instead of $U^{}_{\mu i} = U^{}_{\tau i}$
(for $i = 1,2,3$) \cite{Xing:2014zka}, and the possibility of a flavor democracy
for the first column of $U^{}_{\rm PMNS}$ (i.e.,
$U^{}_{e 1} = U^{}_{\mu 1} = U^{}_{\tau 1}$) has been discarded from the beginning
since it is definitely incompatible with current neutrino oscillation data
\cite{Zyla:2020zbs}.

At this point it is worth emphasizing that the translational symmetry
of ${\cal L}^{}_{\rm M}$ is useful in the following two aspects: on the one hand,
it forces the smallest neutrino mass to be zero; on the other hand, it helps
to partly constrain the texture of $M^{}_\nu$ and the pattern of $U^{}_{\rm PMNS}$.
Nevertheless, such constraints cannot be explicitly achieved unless the coefficients
of $z^{}_\nu$ in the translational transformation of $\nu^{}_{\alpha \rm L}$ in
Eq.~(7) are specified, as we have already seen from the example taken in Eqs.~(10)
and (11). Note that the model building exercises based on most of the discrete flavor
symmetries actually face a similar problem \cite{Xing:2019vks,Feruglio:2019ybq}:
one usually has to follow a somewhat contrived way to assign the irreducible
representations of a flavor symmetry group to the relevant fermion and scalar
fields in a concrete model, and the primary guideline in this connection is
fully empirical or phenomenological --- just to fit the available experimental
data as well as possible. The same is true of the charged fermion sector,
unfortunately, as one will see later on. Before a convincing breakthrough is
made in flavor dynamics, we find it useful to explore all the possibilities
from the bottom up.

Provided $M^{}_l$ is neither diagonal nor Hermitian, one can diagonalize
$M^{}_l M^\dagger_l$ with the help of a unitary transformation
$U^\dagger_l M^{}_l M^\dagger_l U^{}_l = {\rm diag}\{m^2_e, m^2_\mu, m^2_\tau\}$. In
this case the PMNS lepton flavor mixing matrix is expressed as
$U^{}_{\rm PMNS} = U^\dagger_l U^{}_\nu$, where $U^{}_\nu$ is equivalent
to the unitary matrix $U$ used to diagonalize $M^{}_\nu$ in the above discussions.
Of course, the mass term of the charged leptons or Dirac neutrinos may also possess
a possible translational flavor symmetry of this kind, so may the mass term of
the up- or down-type quarks.

\section{Dirac fermions}

We proceed to discuss the Dirac fermion mass terms in an opposite way.
If the massive neutrinos are of the Dirac nature, it is possible to treat them on the
same footing as the charged leptons and quarks. Without loss of any generality, a
Dirac fermion mass matrix $M^{}_\psi$ can always be taken to be Hermitian
after a proper choice of the flavor basis in the SM or its extensions which
have no flavor-changing right-handed currents \cite{Frampton:1985qk}. In
this case one may diagonalize $M^{}_\psi$ via the unitary transformation
$V^\dagger M^{}_\psi V = {\rm diag}\{\lambda^{}_1 , \lambda^{}_2 ,
\lambda^{}_3\}$ with $\lambda^{}_i$ being the eigenvalues of $M^{}_\psi$
(i.e., $m^{}_i = |\lambda^{}_i|$ are physical masses of the charged or neutral
fermions under consideration, either for leptons or for quarks).
The effective mass term for a given category of the fundamental fermions with
the same electric charge can be written in the chosen basis as follows:
\begin{eqnarray}
-{\cal L}^{}_{\rm D} = \sum_\alpha\sum_\beta \left[\overline{\psi^{}_{\alpha \rm L}}
\hspace{0.05cm} \langle m\rangle^{}_{\alpha\beta} \hspace{0.05cm}
\psi^{}_{\beta \rm R} \right] + {\rm h.c.} \; ,
\end{eqnarray}
in which the subscripts $\alpha$ and $\beta$ run over the flavor indices of the
Dirac neutrinos, charged leptons, up-type quarks or dow-type quarks, and
\begin{eqnarray}
\langle m\rangle^{}_{\alpha\beta} = \langle m\rangle^{*}_{\beta\alpha}
= \sum_i \left(\lambda^{}_i V^{}_{\alpha i} V^*_{\beta i} \right) \;
\end{eqnarray}
holds thanks to the Hermiticity of $M^{}_\psi$.
Now let us require that ${\cal L}^{}_{\rm D}$ keep unchanged under a translational
transformation of the left- and right-handed fermion fields in the flavor space,
\begin{eqnarray}
\psi^{}_{\alpha \rm L(R)} \to \psi^{}_{\alpha \rm L(R)}
+ n^{}_\alpha z^{}_{\psi \rm L(R)} \; ,
\end{eqnarray}
where $n^{}_\alpha$ and $z^{}_{\psi \rm L(R)}$ are the flavor-dependent complex
numbers and a constant spinor field anticommuting with the fermion fields
\footnote{If $z^{}_{\psi \rm L(R)}$ is spacetime-dependent, one should also take into
account the relevant kinetic energy term. That would make things more complicated.
But one example of this kind for the right-handed neutrinos, which are the gauge
$\rm SU(2)^{}_{\rm L}$ singlets, has been discussed in Ref.~\cite{He:2009pt}.},
respectively. We find that ${\cal L}^{}_{\rm D}$ will be invariant with respect to
the transformation made in Eq.~(14) if and only if the conditions
\begin{eqnarray}
&& \sum_\alpha \left[n^{*}_\alpha \langle m\rangle^{}_{\alpha\beta} \right]
= 0 \; ,
\nonumber \\
&& \sum_\beta \left[\langle m\rangle^{}_{\alpha\beta} \hspace{0.05cm}
n^{}_\beta \right] = 0 \; ,
\nonumber \\
&& \sum_\alpha \sum_\beta \left[n^{*}_\alpha
\langle m\rangle^{}_{\alpha\beta} \hspace{0.05cm} n^{}_\beta \right] = 0 \;
\end{eqnarray}
are satisfied. Given these constraints, it is easy to show that the determinant of
$M^{}_\psi$ vanishes, indicating that one of the three fermion masses
$m^{}_i$ (for $i=1,2,3$) must be exactly vanishing. Substituting Eq.~(13)
into Eq.~(15), we immediately obtain
\begin{eqnarray}
&& \sum_i \left[ \lambda^{}_i \sum_\alpha \left(n^{*}_\alpha V^{}_{\alpha i}
\right) V^*_{\beta i} \right] = 0 \; ,
\nonumber \\
&& \sum^{}_i \left[\lambda^{}_i V^{}_{\alpha i} \sum_\beta \left(
V^*_{\beta i} \hspace{0.05cm} n^{}_\beta \right) \right] = 0 \; ,
\nonumber \\
&& \sum^{}_i \left[\lambda^{}_i \sum_\alpha \left(n^{*}_\alpha V^{}_{\alpha i}\right)
\sum_\beta \left(V^*_{\beta i} \hspace{0.05cm} n^{}_\beta \right) \right] = 0 \; .
\end{eqnarray}
If the sum of $n^{*}_\alpha V^{}_{\alpha i}$ over $\alpha$ (or equivalently,
the sum of $V^*_{\beta i} \hspace{0.05cm} n^{}_\beta$ over $\beta$) in
Eq.~(16) were nonzero,
one would be left with an interesting but phenomenologically-disfavored relationship
$V^{}_{\beta i}/V^{}_{\beta j} = {\it constant}$ for $\beta$ taking different flavors
in connection with $m^{}_k = 0$, where $i$, $j$ and $k$ run cyclically over $1$, $2$
and $3$. So this nontrivial solution has to be discarded.

It is therefore straightforward to obtain the other nontrivial solution to Eq.~(16):
\begin{eqnarray}
\sum_\alpha \left(n^{*}_\alpha V^{}_{\alpha i}
\right) = \sum_\beta \left(
V^*_{\beta i} \hspace{0.05cm} n^{}_\beta \right) = 0 \; ,
\end{eqnarray}
in correspondence to $\lambda^{}_i \neq 0$. One can see that
$n^{}_\alpha \propto V^{}_{\alpha j}$ and $n^{}_\beta \propto V^{}_{\beta j}$
with $j \neq i$ satisfy Eq.~(17). Imposing the normalization condition on
$n^{}_\alpha$, we simply take $n^{}_\alpha = V^{}_{\alpha j}$ and
$n^{}_\beta = V^{}_{\beta j}$ associated
with the $j$-th mass eigenvalue $\lambda^{}_j$, and substitute them into
Eq.~(16). We are then left with the consistent results
\begin{eqnarray}
\lambda^{}_j V^*_{\beta j} = 0 \; , \quad
\lambda^{}_j V^*_{\alpha j} = 0 \; , \quad
\lambda^{}_j = 0 \; .
\end{eqnarray}
In other words, the invariance of a Dirac fermion mass term ${\cal L}^{}_{\rm D}$
under the translational transformation made in Eq.~(14) implies that the three
flavor-dependent complex numbers $n^{}_\alpha$ can be fully determined by the
elements of $V$ in its $j$-th column corresponding to $\lambda^{}_j = 0$.

The reasoning made above for the Dirac fermions can also be extended to the Majorana
neutrinos, or vice versa. While the possibility of $m^{}_1 = 0$ (or $m^{}_3 = 0$) is
still allowed by current experimental data, none of $m^{}_e = 0$, $m^{}_u = 0$ and
$m^{}_d = 0$ are true in nature. But the observed striking hierarchies
$m^{}_e \ll m^{}_\mu \ll m^{}_\tau$, $m^{}_u \ll m^{}_c \ll m^{}_t$ and
$m^{}_d \ll m^{}_s \ll m^{}_b$ \cite{Zyla:2020zbs} indicate that the zero mass limit
for the first-family charged fermions is actually a reasonable starting point for model
building, and the nonzero but small values of $m^{}_e$, $m^{}_u$ and $m^{}_d$ can be
naturally attributed to either the tree-level perturbations \cite{Fritzsch:1977za,Weinberg:1977hb,Wilczek:1977uh} or the loop-level corrections
\cite{Weinberg:1972ws,Weinberg:2020zba}.

Note that $m^{}_u = 0$ used to be the most economical solution to the
strong CP problem in quantum chromodynamics (QCD) \cite{Weinberg:1977ma,Wilczek:1977pj},
but it has been discarded today. In any case $m^{}_u =0$ can be regarded
as a straightforward consequence of the translational symmetry of ${\cal L}^{}_{\rm D}$
for the up-type quarks as discussed above, and it is well in tune with the naturalness
principle advocated by 't Hooft \cite{tHooft:1979rat}. The finite
values of $m^{}_u$, $m^{}_d$, $m^{}_e$ and $m^{}_1$ (or $m^{}_3$) can therefore be
generated from some slight breaking of such an unconventional flavor symmetry.
To keep the flavor mixing pattern obtained in the zero mass limit
unspoiled, the simplest phenomenological way to break the translational symmetry of
${\cal L}^{}_{\rm D}$ (or ${\cal L}^{}_{\rm M}$) is just to add a diagonal and
flavor-universal mass term \cite{Friedberg:2006it,Friedberg:2007ba,Friedberg:2007uk}.
Namely, ${\cal L}^{}_{\rm D}$ in Eq.~(12) can now be written as follows
\footnote{A similar term of the form
$\displaystyle \frac{1}{2} m^{}_0 \sum_\alpha \left[\overline{\nu^{}_{\alpha \rm L}}
\hspace{0.02cm} \left(\nu^{}_{\alpha \rm L}\right)^c \right]$ can be added to
${\cal L}^{}_{\rm M}$ in Eq.~(4), where $m^{}_0$ characterizes a slight breaking
of the translational symmetry of ${\cal L}^{}_{\rm M}$ and its magnitude is comparable
with the smallest neutrino mass $m^{}_1$ (or $m^{}_3$).}:
\begin{eqnarray}
-{\cal L}^{}_{\rm D} = \sum_\alpha\sum_\beta \left[\overline{\psi^{}_{\alpha \rm L}}
\hspace{0.05cm} \langle m\rangle^{}_{\alpha\beta} \hspace{0.05cm}
\psi^{}_{\beta \rm R} \right] + m^{}_0
\sum_\alpha \left[\overline{\psi^{}_{\alpha \rm L}}
\hspace{0.08cm} \psi^{}_{\alpha \rm R} \right] + {\rm h.c.} \; ,
\end{eqnarray}
where $m^{}_0$ measures the explicit symmetry breaking effect and its magnitude
is expected to be comparable with the mass of the electron, the up quark or
the down quark. In this case the smallness of $m^{}_0$ implies that ${\cal L}^{}_{\rm D}$
may still have an approximate translational symmetry.

To illustrate, let us consider the up- and down-quark sectors in the
$m^{}_u = m^{}_d = 0$ limits by simply taking
\begin{eqnarray}
V^{}_{\rm q} = \frac{1}{\sqrt 6} \left(\begin{matrix} 2 & \sqrt{2} & 0 \cr
-1 & \sqrt{2} & -\sqrt{3} \cr
-1 & \sqrt{2} & \sqrt{3} \cr \end{matrix} \right)
\left(\begin{matrix} 1 & 0 & 0 \cr
0 & c^{}_{\rm q} & s^{}_{\rm q} \cr
0 & -s^{}_{\rm q} & c^{}_{\rm q} \cr \end{matrix} \right) \; ,
\end{eqnarray}
where $c^{}_{\rm q} \equiv \cos\vartheta^{}_{\rm q}$ and
$s^{}_{\rm q} \equiv \sin\vartheta^{}_{\rm q}$ (for $\rm q = u$ or $\rm d$).
In this case the CKM flavor mixing matrix turns out to be
\begin{eqnarray}
V^{}_{\rm CKM} = V^\dagger_{\rm u} V^{}_{\rm d} =
\left(\begin{matrix} 1 & 0 & 0 \cr
0 & \cos\vartheta & \sin\vartheta \cr
0 & -\sin\vartheta & \cos\vartheta \cr \end{matrix} \right) \; ,
\end{eqnarray}
where $\vartheta = \vartheta^{}_{\rm d} - \vartheta^{}_{\rm u}$ is a
nontrivial but small quark mixing angle. Note that
the structural parallelism between $V^{}_{\rm u}$
and $V^{}_{\rm d}$ (or equivalently, between $M^{}_{\rm u}$ and $M^{}_{\rm d}$)
assures that the resulting $V^{}_{\rm CKM}$ should not deviate far from
the identity matrix
\footnote{In comparison, the charged-lepton and neutrino
mass matrices must be quite different in their textures so as to give rise to
significant effects of lepton flavor mixing.}.
Combining Eqs.~(13) and (21), we find that the reconstructed
up-type quark mass matrix $M^{}_{\rm u}$ has the same form as $M^{}_\nu$
in Eq.~(11) with the replacements
$a \equiv \left( m^{}_c c^2_{\rm u} + m^{}_t s^{2}_{\rm u} \right)/3$,
$b \equiv \left( m^{}_c s^{2}_{\rm u} + m^{}_t c^2_{\rm u} \right)/2$
and $c \equiv \left( m^{}_c - m^{}_t \right) c^{}_{\rm u} s^{}_{\rm u}/\sqrt{6}$,
so does the down-type quark mass matrix $M^{}_{\rm d}$ with the corresponding
replacements. To generate the masses for the first-family quarks together with
the other two flavor mixing angles and CP violation, one needs to properly break
the translational symmetry in the up- and down-quark sectors.

If the translational symmetry of a Dirac or Majorana mass term is obtained
at a superhigh energy scale, it may also be broken at the electroweak scale
due to the quantum corrections described by the renormalization-group
equations (RGEs) \cite{Ohlsson:2013xva}. But it has been shown that a nonzero 
value of $m^{}_1$ (or $m^{}_3$) of ${\cal O}(10^{-13}) ~{\rm eV}$ cannot 
be generated from $m^{}_1 =0$ (or $m^{}_3 =0$) unless the two-loop RGE running 
effects are taken into account for the Majorana neutrinos
\cite{Davidson:2006tg,Xing:2020ezi,Zhou:2021bqs}, and
such a tiny mass and the associated Majorana CP phase do not play any
seeable role in neutrino physics. The masses of the first-family Dirac fermions
are in general insensitive to the two-loop RGE-induced quantum corrections either
\cite{Barger:1992ac,Barger:1992pk,Luo:2002ey}.

\section{Summary}

It has been a common belief in particle physics that behind different families
of the fundamental fermions is some kind of flavor symmetry which can help
understand the salient features of fermion mass spectra and flavor mixing patterns.
Although the origin of the massive Majorana neutrinos may be quite different
from that of the charged leptons and quarks, their flavor structures are
likely to share the same symmetry. In this connection many flavor symmetries
(either Abelian or non-Abelian, either continuous or discrete, either local or
global) have been tried in the past decades, but new ideas are always called for
in order to pin down the true flavor dynamics. That is why we highlight a
possible translational flavor symmetry associated with the mass terms of fundamental
Dirac or Majorana fermions in this work. In particular, we point out that the
translational transformation $\nu^{}_i \to \nu^{}_i + z^{}_\nu$ for a massless neutrino
field $\nu^{}_i$ is equivalent to the translational
transformation $\nu^{}_\alpha \to \nu^{}_\alpha + U^{}_{\alpha i} z^{}_\nu$
in the flavor space with $U^{}_{\alpha i}$ being the
corresponding neutrino flavor mixing matrix elements and $z^{}_\nu$ denoting
a constant spinor field anticommuting with the neutrino fields.

If the effective mass term for a category of Dirac or Majorana
fermions of the same electric charge keeps unchanged under the transformation
$\psi^{}_{\alpha \rm L (R)} \to \psi^{}_{\alpha \rm L (R)} + n^{}_{\alpha}
z^{}_{\psi \rm L(R)}$ in the flavor space, where $n^{}_\alpha$
and $z^{}_{\psi \rm L(R)}$ stand respectively
for the flavor-dependent complex numbers and a constant spinor field anticommuting
with the fermion fields, we have shown that $n^{}_\alpha$ can be
simply identified as the elements $U^{}_{\alpha i}$ in the $i$-th column of the
unitary matrix $U$ used to diagonalize the corresponding Hermitian or symmetric
fermion mass matrix $M^{}_\psi$, and $m^{}_i = 0$ holds accordingly. The reverse is
also true, and thus this translational flavor symmetry can constrain both the fermion
mass spectrum and the flavor mixing pattern. Given the very facts that all the
fundamental fermions of the same nonzero electric charge have a strong mass hierarchy
and current experimental data allow the lightest neutrino to be (almost)
massless, we expect that the zero mass limit for the first-family fermions and the
flavor symmetry behind it should be a natural starting point for model
building and may even help shed light on the secret of flavor dynamics.

\acknowledgements{
The author is indebted to Ferruccio Feruglio, Harald Fritzsch, Yu-Feng Li, 
Shu Luo, Zhen-hua Zhao and Shun Zhou for helpful discussions. This work was 
supported in part by the National Natural Science Foundation of China under 
grants No. 12075254, No. 11775231 and No. 11835013.}


\end{document}